\documentclass[a4paper,12pt]{article}
\usepackage{graphicx, rotating,amssymb}

\ifx\pdfoutput\undefined
\usepackage[dvips,bookmarks]{hyperref}	
\else
\usepackage{hyperref}	
\fi
\hypersetup{colorlinks,bookmarksopen,bookmarksnumbered,citecolor=verdes,
linkcolor=blus,pdfstartview=FitH,urlcolor=rossos}
\def\myurl#1#2{\href{http://#1}{#2}}
\def\hhref#1{\href{http://arxiv.org/abs/#1}{#1}} 

\usepackage{multicol}
\usepackage{color}
\definecolor{rosso}{cmyk}{0,1,1,0.4}
\definecolor{rossos}{cmyk}{0,1,1,0.55}
\definecolor{rossoc}{cmyk}{0,1,1,0.2}
\definecolor{blu}{cmyk}{1,1,0,0.3}
\definecolor{blus}{cmyk}{1,1,0,0.6}
\definecolor{bluc}{cmyk}{1,1,0,0.1}
\definecolor{verde}{cmyk}{0.92,0,0.59,0.25}
\definecolor{verdec}{cmyk}{0.92,0,0.59,0.15}
\definecolor{verdes}{cmyk}{0.92,0,0.59,0.4}

\font\tenrsfs=rsfs10 at 12pt
\font\sevenrsfs=rsfs7
\font\fiversfs=rsfs5
\newfam\rsfsfam
\textfont\rsfsfam=\tenrsfs
\scriptfont\rsfsfam=\sevenrsfs
\scriptscriptfont\rsfsfam=\fiversfs
\def\mathscr#1{{\fam\rsfsfam\relax#1}}

\newcommand{\fig}[1]{~\ref{fig:#1}}
\newcommand{\riga}[1]{\noalign{\hbox{\parbox{\textwidth}{#1}}}\nonumber}

\newcommand{\eq}[1]{~{\rm (\ref{eq:#1})}}

\newcommand{\GeV}{\,{\rm GeV}}
\newcommand{\TeV}{\,{\rm TeV}}

\def\circa#1{\,\raise.3ex\hbox{$#1$\kern-.75em\lower1ex\hbox{$\sim$}}\,}

\newcommand{\DM}{{\rm DM}}

\newcommand{\beq}{\begin{equation}}
\newcommand{\eeq}{\end{equation}}

\def\circa#1{\,\raise.3ex\hbox{$#1$\kern-.75em\lower1ex\hbox{$\sim$}}\,}
\makeatletter

\def\art{\@ifnextchar[{\eart}{\oart}}
\def\eart[#1]#2#3#4#5#6{{\rm #2}, {#3 #4} {\rm (#6) #5} [{\hhref{#1}}]}
\def\hepart[#1]#2{{\rm #2, \hhref{#1}}}
\newcommand{\oart}[5]{{\rm #1}, {#2 #3} {\rm (#5) #4}}

\newcounter{alphaequation}[equation]
\def\thealphaequation{\theequation\hbox to
0.6em{\hfil\alph{alphaequation}\hfil}}
\def\eqnsystem#1{
\def\@eqnnum{{\rm (\thealphaequation)}}
\def\@@eqncr{\let\@tempa\relax \ifcase\@eqcnt \def\@tempa{& & &} \or
  \def\@tempa{& &}\or \def\@tempa{&}\fi\@tempa
  \if@eqnsw\@eqnnum\refstepcounter{alphaequation}\fi
\global\@eqnswtrue\global\@eqcnt=0\cr}
\refstepcounter{equation} \let\@currentlabel\theequation \def\@tempb{#1}
\ifx\@tempb\empty\else\label{#1}\fi
\refstepcounter{alphaequation}
\let\@currentlabel\thealphaequation
\global\@eqnswtrue\global\@eqcnt=0 \tabskip\@centering\let\\=\@eqncr
$$\halign to \displaywidth\bgroup \@eqnsel\hskip\@centering
$\displaystyle\tabskip\z@{##}$&\global\@eqcnt\@ne
\hskip2\arraycolsep\hfil${##}$\hfil& \global\@eqcnt\tw@\hskip2\arraycolsep
$\displaystyle\tabskip\z@{##}$\hfil
\tabskip\@centering&\llap{##}\tabskip\z@\cr}
\def\endeqnsystem{\@@eqncr\egroup$$\global\@ignoretrue} \makeatother

\newcommand{\SU}{\rm SU}

\begin{document}
\begin{center}
{IFUP--TH/2008-05}
{ \hfill SACLAY--T08/034}
\color{black}
\vspace{0.3cm}

{\Huge\bf\color{rossos} Minimal Dark Matter\\ predictions for galactic\\[1.5mm] positrons, anti-protons, photons}

\medskip
\bigskip\color{black}\vspace{0.6cm}

{
{\large\bf Marco Cirelli}$^a$,
{\large\bf Roberto Franceschini}$^b$,
{\large\bf Alessandro Strumia}$^c$
}
\\[7mm]
{\it $^a$ Institut de Physique Th\'eorique, CEA-Saclay \& CNRS, France}\ \footnote{CEA, DSM, Institut de Physique Th\'orique, IPhT, CNRS, MPPU, URA2306, Saclay, F-91191 Gif-sur-Yvette, France}\\[3mm]
{\it $^b$ Scuola Normale Superiore \& INFN, Pisa, Italy}\\[3mm]
{\it $^c$ Dipartimento di Fisica dell'Universit{\`a} di Pisa \& INFN, Italia}
\end{center}

\bigskip

\centerline{\large\bf\color{blus} Abstract}
\begin{quote}
\color{black}\large
We present the energy spectra of the fluxes of positrons, anti-protons and photons
generated by Dark Matter annihilations in our galaxy,
as univocally predicted by the model of Minimal Dark Matter. Due to multi-TeV masses and to the Sommerfeld enhancement of the annihilation cross section, distinctive signals can be generated above the background, even with a modest astrophysical boost factor, in the range of energies soon to be explored by cosmic ray experiments.

\end{quote}




\section{Introduction}
We consider Minimal Dark Matter~\cite{MDM,MDM2} (MDM), i.e.\ we assume that the DM 
is the neutral component of one single weak multiplet, that interacts with SM particles only via (broken) gauge SM interactions. The assignment of spin and 
$\SU(2)_L \times {\rm U(1)}_Y$ quantum numbers fully identifies each different MDM candidate:
the full list, together with a short list of the most interesting candidates, has been presented in~\cite{MDM,MDM2}. 

The main virtues of such model (which is not inspired by more ambitious beyond-the-SM constructions like super symmetry or extra dimensions) can therefore be summarized in terms of economy and predictiveness. 
The model has no free parameters as all DM couplings are predicted by gauge invariance and the DM 
mass is determined by matching the relic abundance, $\Omega_{\rm DM} h^2 =0.110 \pm 0.005$~\cite{cosmoDM}.
A particularly interesting MDM candidate is the fermionic $\SU(2)_L$ 5-plet with hypercharge $Y=0$, that is automatically stable
on cosmological time-scale thanks to the SM gauge and Lorentz symmetries, without having to impose 
ad-hoc parities (like $R$-parity or KK-parity). 
We will restrict our study to three particularly interesting MDM candidates: the fermionic quintuplet already mentioned above; the fermionic 3-plet with hypercharge $Y=0$ (the MDM candidate that has the same quantum numbers of the supersymmetric `wino');
the scalar triplet with $Y=0$.

\medskip

We here compute the ``indirect DM signals'', generated by DM DM
annihilations into  $\bar p$, $e^+$, $\gamma$ in our galaxy~\cite{pioneers}.
Unlike DM candidates motivated by a successful natural
solution to the hierarchy problem (which should therefore have mass below or around the $Z$ mass), MDM predicts specific multi-TeV values for the DM mass $M_{\rm DM}$, and annihilation cross sections enhanced by electroweak Sommerfeld corrections: these two features imply a distinctive DM signal that can be tested by running and future experiments like PAMELA~\cite{PAMELA} and AMS-02~\cite{AMS02}, dedicated to extending our knowledge of  galactic Cosmic Ray (CR) spectra up to higher energies.

\begin{figure}[t]
\begin{center}
\includegraphics[width=0.4\textwidth]{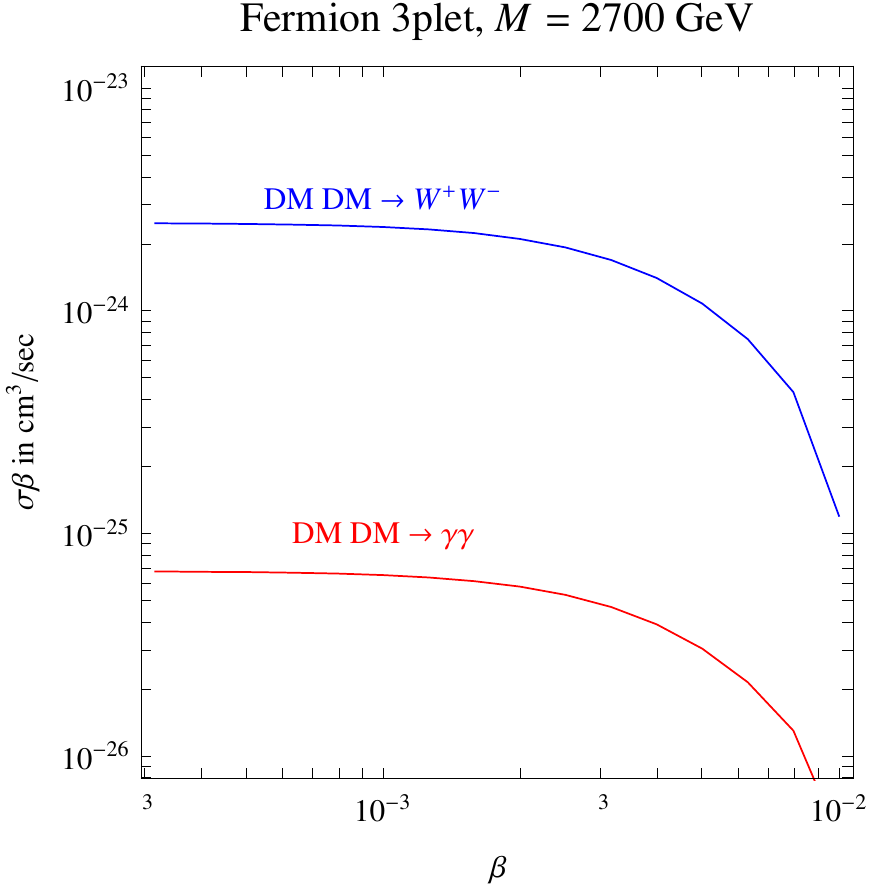}\qquad
\includegraphics[width=0.4\textwidth]{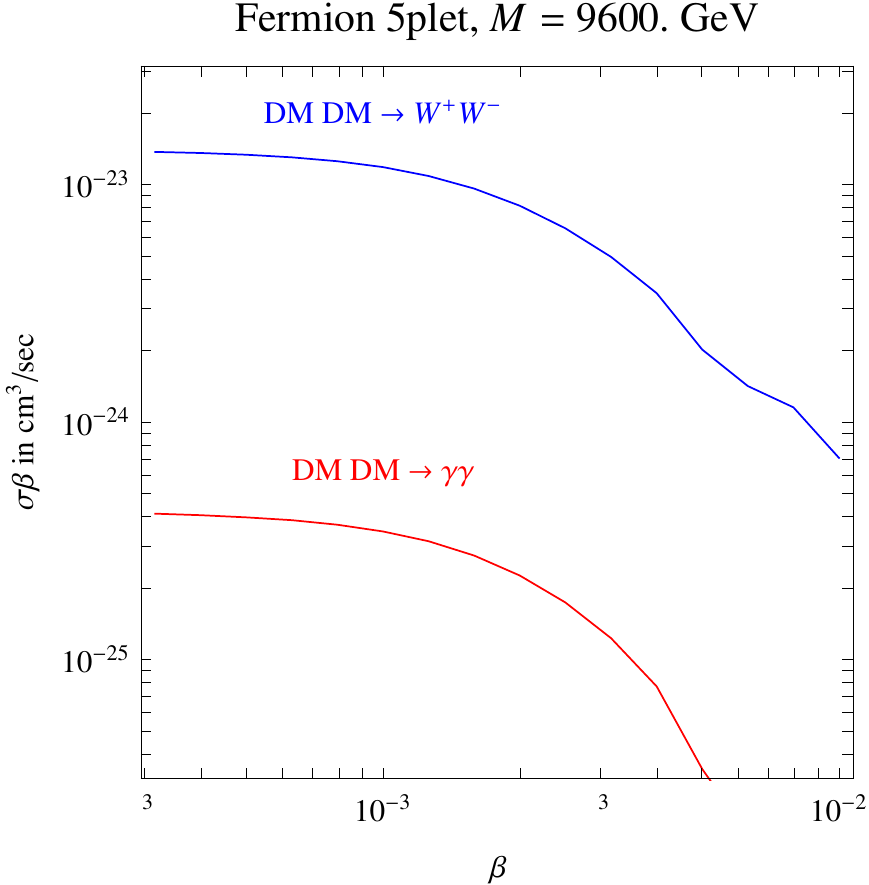}
\caption{\em\label{fig:sigmav} Velocity dependence of Sommerfeld-enhanced MDM annihilation cross sections, for the two candidates that we mainly consider.}
\end{center}
\end{figure}

\section{Energy spectra at production}

MDM annihilates at tree level into $W^+W^-$, and at loop level into
$\gamma\gamma$, $\gamma Z$, $ZZ$.
The relative cross-sections are significantly affected by non-perturbative Sommerfeld corrections~\cite{Hisano}, and we use the results of~\cite{MDM2}. 
As a consequence of Sommerfeld corrections, the DM DM annihilation cross sections exhibit a quite steep dependence on $M_\DM$ 
and can vary by about one order of magnitude
within the range allowed at $3\sigma$ by the cosmological DM abundance as computed assuming thermal freeze-out (see e.g.\ fig.s 2--5 in~\cite{MDM2}).
For the same reason, the cross section $\sigma v$ also depends on the DM velocity $v$, reaching a maximal value for $v\to 0$, as shown in fig.\fig{sigmav}. The average DM velocity in our galaxy, $v\approx 10^{-3}$, is however low enough that $\sigma v$ is close to its maximal value, which we assume.

For definiteness, in the following we adopt the following best-fit values of $M_\DM$ and $\langle \sigma v \rangle$ for the MDM candidates that we consider: 
\begin{eqnsystem}{sys:sample}
M_\DM=9.6\TeV,\qquad&& \langle \sigma v \rangle_{WW} = 1.1\cdot 10^{-23} {{\rm cm}^3\over {\rm sec}},\qquad
\langle \sigma v \rangle_{\gamma\gamma}= 3\cdot 10^{-25} {{\rm cm}^3\over {\rm sec}}\\[1mm]
\riga{for the fermion quintuplet with $Y=0$, and}\\[-4mm]
M_\DM=2.7\TeV,\qquad&& \langle \sigma v \rangle_{WW}= 0.21\cdot 10^{-23} {{\rm cm}^3\over {\rm sec}},\qquad
\langle \sigma v \rangle_{\gamma\gamma}= 0.58\cdot 10^{-25} {{\rm cm}^3\over {\rm sec}}\\[1mm]
\riga{for the fermion triplet with $Y=0$, and}\\[-4mm]
M_\DM=2.5\TeV,\qquad&& \langle \sigma v \rangle_{WW}= 3.6\cdot 10^{-23} {{\rm cm}^3\over {\rm sec}},\qquad
\langle \sigma v \rangle_{\gamma\gamma}= 9.4\cdot 10^{-25} {{\rm cm}^3\over {\rm sec}}
\end{eqnsystem}
 for the scalar triplet with $Y=0$.  We will not plot predictions to the scalar triplet,
 that can be easily read out from the corresponding predictions for the fermion triplet,
 taking into account that they have a similar mass, and multiplying all rates by a factor of about 16~\cite{MDM2},
due to a large Sommerfeld enhancement.
The other automatically stable MDM candidate, the scalar eptaplet, is expected to have a mass $M_\DM \sim 25\TeV$,
but we cannot reliably predict its annihilation cross sections. 

Annihilation cross sections into $\gamma Z$ and $ZZ$ are given by
\beq \sigma_{\gamma Z} = 2\sigma_{\gamma\gamma}/\tan^2\theta_{\rm W}=
6.5 \sigma_{\gamma\gamma}
,\qquad
\sigma_{ZZ} = \sigma_{\gamma\gamma}/\tan^4\theta_{\rm W} = 10.8\sigma_{\gamma\gamma}.\eeq
for all MDM candidates with $Y=0$.

\begin{table}\footnotesize
$$
\begin{array}{rcl|ccccccccc}
\multicolumn{3}{c}{\rm Process}& c_0 & c_1 & c_2 & c_3 & c_4 & c_5 & c_6 & c_7 &c_8\\  \hline
WW&\to& e^+ & -1.895 & 2.821 & 6.299 & 7.563 & 6.914 & 4.812 & 2.367 \
& 0.7273 & 0.1050\\ 
WW&\to& \bar{p} & -12.26 & -18.84 & -29.21 & -36.62 & -33.63 & -20.98 \
& -8.006 & -1.411 & 0\\ 
WW&\to& \gamma & -6.751 & -5.741 & -3.514 & -1.964 & -0.8783 & \
-0.2512 & -0.03369 & 0 & 0\\ 
ZZ&\to& e^+ & -2.485 & 2.809 & 5.501 & 4.901 & 2.953 & 1.252 & 0.3424 \
& 0.04574 & 0\\ 
ZZ&\to& \bar{p} & -8.423 & -8.396 & -9.168 & -8.652 & -5.549 & -1.591 \
& 0.3074 & 0.2717 & 0\\ 
ZZ&\to& \gamma & -7.418 & -6.829 & -5.308 & -4.105 & -2.601 & -1.164 \
& -0.3256 & -0.04312 & 0\\ 
%
\end{array}$$
\caption{\em\label{tab:cn}Coefficients for the analytic approximation in eq.\eq{app} to the fragmentation functions.}
\end{table}


\bigskip

We next need to compute the energy spectra of $e^+,p^-,\gamma$ produced by decays of SM vectors.
Instead of using the results available from  the literature, we performed an independent computation.
Indeed, to our knowledge, previous `decay' computations do not take into account spin correlations of
SM vectors $V$ in the intermediate state.
We instead compute the full matrix element for $\DM\, \DM \to V\bar V \to 4$ fermions.
For example, including  spin correlations the energy spectra of primary positrons directly produced
in the annihilation $\DM\, \DM \to W^+ W^- \to $ 4 fermions, is
\beq \left.\frac{dN_{e^+}}{dx}\right|_{\rm primary} = \frac{1-2x+2x^2}{6}\qquad \hbox{instead of}\qquad \frac{1}{9}\hbox{ (no spin correlations)}\eeq
where $x = E_{e^+}/M_\DM$ and $M_\DM\gg M_W$.
The same $x$-dependence applies to all massless fermions and SM vectors, and it
arises as follows (see e.g.~\cite{Barger}).  SM vectors are produced isotropically in the $\DM\,\DM$ rest frame,
with equal transversally polarized helicities $h$: both $+$ or both $-$.
This is a characteristic of $s$-wave annihilations, that dominate in the non relativistic limit.
A $V$ at rest decays into massless fermions with angular distribution
\beq \frac{dN}{d\cos\theta} \propto (1 + \cos\theta)^2\eeq
where $\cos\theta$ is the angle between the direction of the fermion and 
the spin of the vector.

\begin{figure}[t]
\begin{center}
\includegraphics[width=\textwidth]{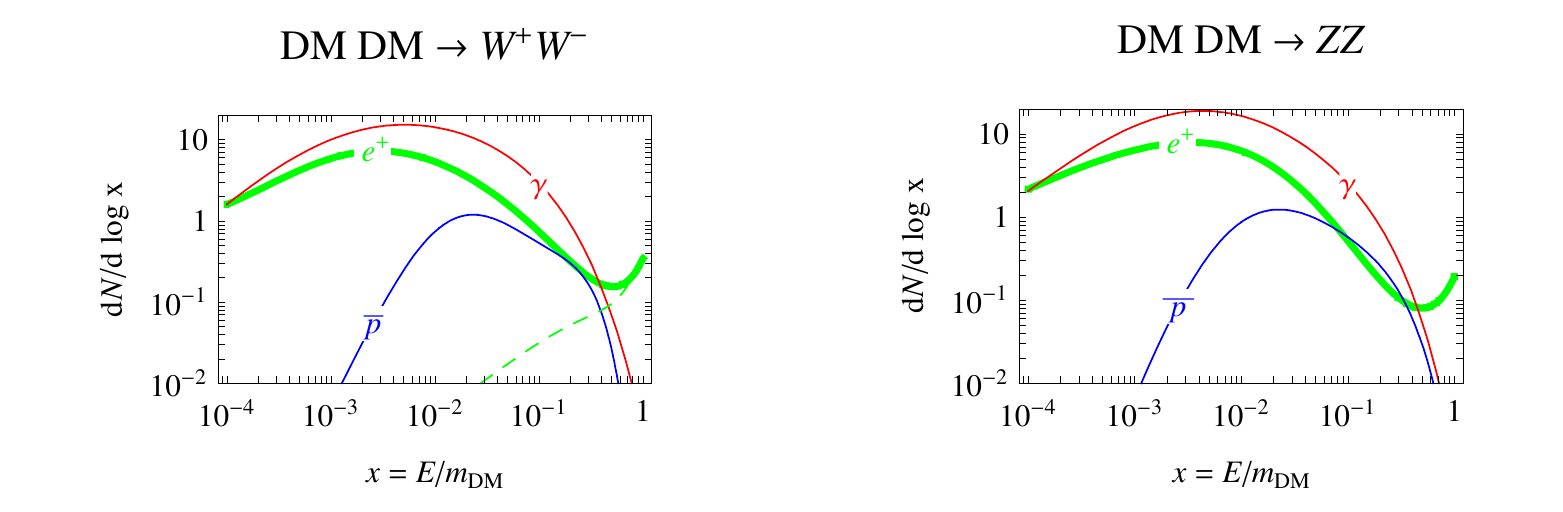}
\caption{\em\label{fig:Fragmentation} Energy spectra of $e^+,\bar p, \gamma$ produced by non-relativistic
$\DM\, \DM$ annihilations into SM vectors.
Only $e^+$ have a secondary component (dashed green line shown on the $W^+W^-$ plot), that dominates at large $x\sim 1$.
}
\end{center}
\end{figure}

We generated 4 fermions events with the   {\sc MadGraph} 4.2~\cite{MadGraph} MonteCarlo event generator,
where we extended the SM to incorporate the DM particle and its charged partner.  This extension incorporates proper tree level interactions among the $W$ and the new particles. The interactions of the new fields with the $Z$ boson arises at one loop, and has been taken into account effectively adding a tree level vertex between the $Z$ and the neutral DM. Although not physical, this vertex is suitable for an easy implementation in the model and at the same time provide realistic $\DM\, \DM \to Z Z$ collisions. In fact we checked that it leads to the expected isotropical production of $Z$ pairs with same helicity, thus proving it to be equivalent to the real one loop vertex.
Actually, in the case of $\DM\, \DM \to W^+ W^- \to $ 4 fermions, we instead used a MonteCarlo routine written by us.

The decay products of the $W$s and the $Z$s produce QED and QCD radiation. This emission is simulated through the parton shower MonteCarlo {\sc Pythia}~8.1~\cite{Pythia}, in which we allowed emission from all the final state particles and resonances. This does not take into account the emission from intermediate states of the process as simulated at the matrix element level. This emission is discussed below.

The formation of hadronic states out of the shower's products has been simulated with {\sc Pythia} as well. Since we are interested to the observation of stable particles we explicitly requested the decay of $\mu^\pm,\,\tau^\pm$ and of all unstable mesons and baryons (including $n$ and $\bar{n}$).

The whole showering, hadronizations and decay process results in the production of  $e^+,p^-,\gamma$ with lower energy $ x\ll 1$.

The final energy spectra are plotted in fig.\fig{Fragmentation} and table~\ref{tab:cn} provides the numerical coefficients $c_n$ in
the analytic approximation
\beq\label{eq:app}
\frac{dN}{d\ln x} = \exp\left[\sum_n \frac{c_n}{n!} \ln^n x\right].\eeq
These results apply in the limit $M_\DM\gg M_{W,Z}$ and have no logarithmic dependence on $M_\DM$, 
since the virtuality of final state particles is $\sim M_{W,Z} \ll M_\DM$.

As {\sc Pythia} only takes into account brehmstrahlung from the final state fermions,
we must separately add the photons produced by brehmstrahlung from $W^\pm$ and from the fermionic charged components $\DM^\pm$ (relevant at $x\circa{<} 1$)~\cite{BergstromBringmann}.
Both particles have virtuality $\sim M_\DM$, leading to a dependence on $\ln\epsilon$, where
$\epsilon=M_V/M_\DM$:
\beq \left.\frac{dN_\gamma}{dx}\right|_{{\rm from~}W^\pm} = \frac{\alpha}{\pi} \frac{1-x}{x} \ln \frac{4(1-x)}{\epsilon^2}\eeq
and~\cite{BergstromBringmann}
\beq \left.\frac{dN_\gamma}{dx}\right|_{{\rm from~}\DM^\pm} = \frac{\alpha}{\pi}\left[\frac{4(1-x+x^2)^2}{x(1-x)}
\ln \frac{2}{\epsilon} +2
\frac{8-3 x^5+16 x^4-37 x^3+42 x^2-24 x}{(1-x) (2-x)^3 x} \ln (1-x)+\right.\eeq
$$\qquad \left.-2\frac{2 x^6-10 x^5+20 x^4-22
   x^3+19 x^2-12 x+4}{(1-x) (2-x)^2 x}\right]$$
where $x\equiv E_\gamma/M_\DM<1-\epsilon$.

We next need to consider how $\gamma,e^+,\bar p$ are produced and propagate in our galaxy.

\begin{figure}[t]
\begin{center}
\includegraphics[width=0.45\textwidth]{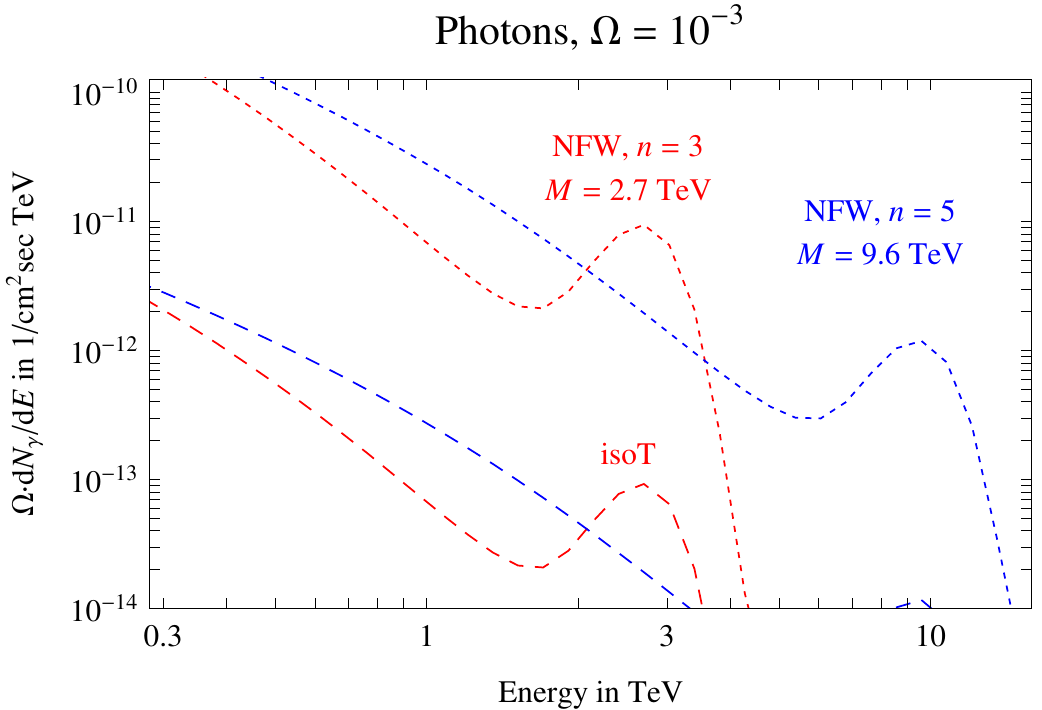}
\caption{\em\label{fig:gamma} {\bf Photon flux} from the galactic center for the isothermal (dashed, $\bar J=13.5$) and NFW (dotted, $\bar J=1380$) DM density profiles.}
\end{center}
\end{figure}


\section{Photons}

\subsection{Astrophysics}
We consider three possible DM halo profiles: cored isothermal~\cite{IsoT}, the Navarro-Frenk-White (NFW)~\cite{NFW} and Moore~\cite{Moore04}.
In all cases the DM density profile can be parameterized as
\beq \rho(r) =
\rho_{\odot} \, \left[{\displaystyle \frac{r_{\odot}}{r}}
\right]^{\gamma} \, \left[ {\displaystyle \frac{1 \, + \, \left(
r_{\odot} / r_{s} \right)^{\alpha}} {1 \, + \, \left( r / r_{s}
\right)^{\alpha}}} \right]^{\left( \beta - \gamma \right) /
\alpha}  \label{eq:rho} \eeq 
where $r_{\odot} = 8.5\,{\rm kpc}$ is the Earth distance from the galactic center,
$\rho_\odot\equiv \rho(r_\odot)$ is the DM density at the Earth position
(we assume $\rho_\odot= 0.3\GeV/{\rm cm}^3$: 
values in the range $0.2-0.7$ are considered in the literature~\cite{Kamionkowski} and the $\gamma,e^+,\bar p$ fluxes scale as $\rho^2$)
and the $\alpha,\beta,\gamma,r_s$ profile parameters are:
$$\begin{tabular}{c|ccccc}
Halo model & $\alpha$ & $\beta$ & $\gamma$ & $r_s$ in kpc \\
\hline
Cored isothermal~\cite{IsoT}&2&2&0&5\\
Navarro, Frenk, White~\cite{NFW} &        1        &        3        &        1        &        20       \\
Moore~\cite{Moore04} &        1      &        3        &        1.16      &        30       
\end{tabular}$$

As well known, the NFW and Moore exhibit a cusp at the center of the galaxy.\footnote{In various numerical computations, it is convenient to
smooth out this behavior adopting the prescription discussed in~\cite{smoothing}. It simply amounts to replace the divergent profile by a well behaved one below an arbitrarily chosen critical radius of $r_{\rm crit} = 0.5\, {\rm kpc}$ from the galactic center, while preserving the absolute number of annihilations in that region. More precisely, we use
$$
\rho(r<r_{\rm crit}) = \rho(r_{\rm crit}) \left[1 + \frac{2 \pi^2}{3} \left( \frac{3}{3-2\gamma} -1 \right) \left( \frac{\sin(\pi r/r_{\rm crit})}{\pi r/r_{\rm crit}} \right)^2 \right]^{1/2}.
\label{eq:smoothing}
$$}

Photons propagate freely.
The differential flux of photons received from a given angular direction $d\Omega$ is
\beq  \frac{d \Phi_\gamma}{d\Omega\,dE} = \frac{1}{2}\frac{c}{4\pi} \frac{\rho_\odot^2}{M_{\rm DM}^2} J
\sum _f\langle \sigma v\rangle_f \frac{dN_\gamma^f}{dE}  ,\qquad
J = \int_{\rm line-of-sight} \frac{ds}{r_\odot} \left(\frac{\rho}{\rho_\odot}\right)^2 
\eeq
where the adimensional quantity $J$ encodes the astrophysical uncertainty.
When observing a region with angular size $\Omega$ 
the factor $J ~d\Omega$ gets replaced by
$ \bar J \Omega = \int_\Omega J~ d\Omega$.
For $\Omega=10^{-3}$ centered around the galactic center one has $\bar J=13.5$ for the isothermal DM profile, $\bar J=1380$ for the NFW profile,
$\bar J=3830$ for the Moore profile.

\subsection{Results}
Fig.\fig{gamma} shows the predicted energy spectrum of the flux of photons produced by MDM annihilations.  
It somewhat differs from the analogous figure in~\cite{MDM2} because we more precisely computed the photon spectrum.
We assumed realistic detector parameters:
an energy resolution of $15\%$,
and that the region observed has angular size $\Omega = 10^{-3}$ 
centered around the galactic center.
It can be rescaled to any other search strategy.
For example, the H.E.S.S.~\cite{HESS} experiment has a much better angular resolution,
that allows it to resolve the black hole at the center of our galaxy.
We do not here address which observational strategy maximizes the sensitivity to MDM photons:
focus on the black hole or subtract it; 
focus on regions of the galaxy far from the center that have less astrophysical $\gamma$ sources
or on nearby galaxies.
These choices only affect $\bar{J}\Omega$ and not the energy spectrum of MDM photons.
The scalar eptaplet with mass $M_{\rm DM}\approx 25\TeV$ provides an energy spectrum that resembles
the one emitted by the galactic center.

\section{Positrons}

\subsection{Astrophysics}

\begin{figure}[t]
\begin{center}
\includegraphics[width=0.45\textwidth]{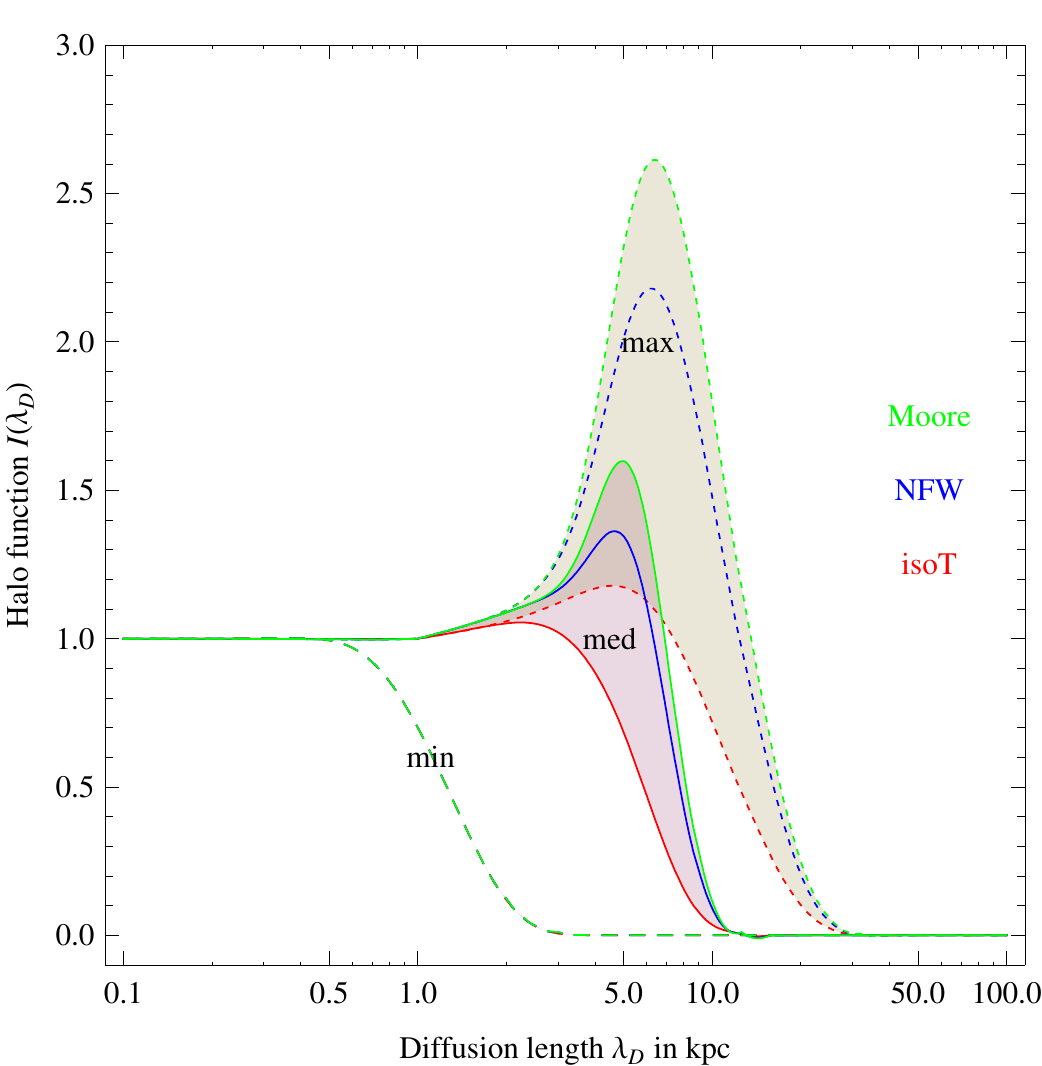}\qquad
\includegraphics[width=0.45\textwidth]{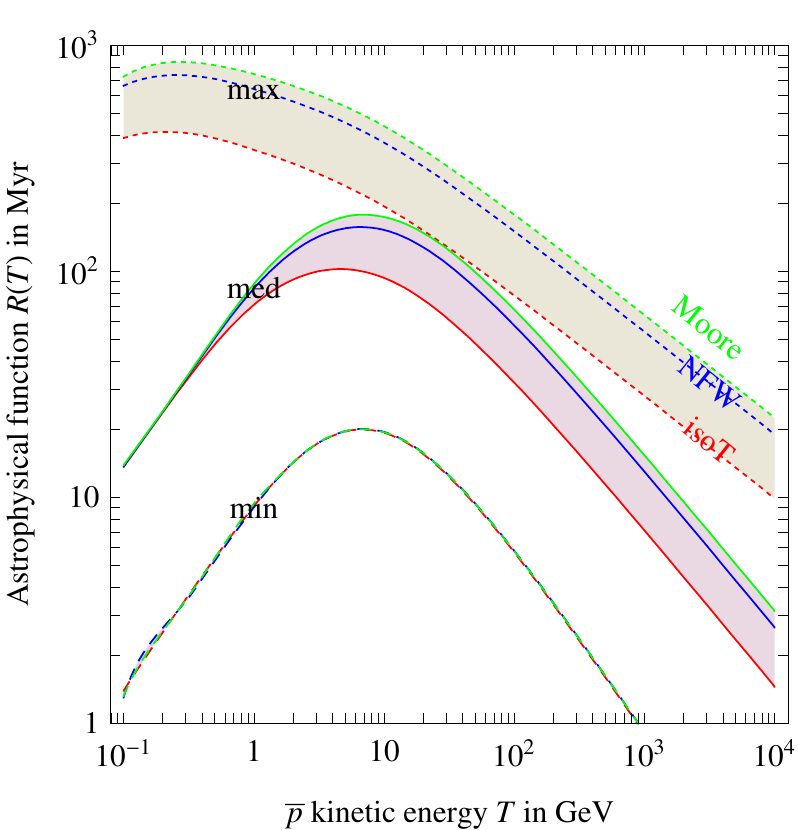}
\caption{\em\label{fig:HaloI} Left: The uncertain `halo function' $I(\lambda_D)$ of eq.\eq{fluxpositrons} that encodes the astrophysics of DM DM annihilations into positrons and their propagation up to the Earth. The diffusion length is related to energy losses as in eq.\eq{lambdaD}.
Right: The $\bar p$ astrophysical function $R(T)$ of eq.\eq{RT}, computed under different assumptions.
In both cases, the dashed (solid) [dotted] bands assumes the {\rm min (med) [max]} propagation configuration of eq.\eq{proparampositrons} and eq.\eq{proparam} respectively.
Each  band contains 3 lines, that correspond to the isothermal (red lower lines), NFW (blue middle lines) and Moore (green upper lines) DM density profiles.
}
\label{default}
\end{center}
\end{figure}

The positron flux per unit energy from DM annihilations in any point in space and time is given by $\Phi_{e^+}(t,\vec x,E) = v_{e^+} f/4\pi$
(units $1/\GeV\cdot{\rm cm}^2\cdot{\rm s}\cdot{\rm sr}$)
where $v_{e^+}$ is the positron velocity (essentially equal to $c$ in the regimes of our interest) and the positron number density per unit energy, $f(t,\vec x,E)= dN_{e^+}/dE$,
obeys the diffusion-loss equation:
\beq \label{eq:diffeq}\frac{\partial f}{\partial t}-
K(E)\cdot \nabla^2f - \frac{\partial}{\partial E}\left( b(E) f \right) = Q\eeq
with diffusion coefficient $K(E)=K_0 (E/\GeV)^\delta$
and energy loss coefficient $b(E)=E^2/(\GeV\cdot \tau_E)$ with $\tau_E = 10^{16}\,{\rm s}$.
They respectively describe transport through the turbulent magnetic fields and energy loss due to
synchrotron radiation and inverse Compton scattering on CMB photons and on infrared galactic starlight.
Eq.~(\ref{eq:diffeq}) is solved in a diffusive region with the shape of a solid flat cylinder that sandwiches the galactic plane, with height $2L$ in the $z$ direction and radius $R=20\,{\rm kpc}$ in the $r$ direction~\cite{DiffusionCylinder}. The location of the solar system corresponds to $\vec x  = (r_{\odot}, z_{\odot}) = (8.5\, {\rm kpc}, 0)$.
The boundary conditions impose that the positron density $f$ vanishes on the surface of the cylinder, outside of which positrons freely propagate and escape.
Values of the propagation parameters $\delta$, $K_0$ and $L$ are deduced from a variety of cosmic ray data and modelizations. We adopt the sets discussed in~\cite{FornengoDec2007}:
\beq \begin{tabular}{lccc}
Model  & $\delta$ & $K_0$ in kpc$^2$/Myr & $L$ in kpc  \\
\hline 
min (M2)  & 0.55 &  0.00595 & 1 \\
med  & 0.70 &  0.0112 & 4  \\
max (M1)  & 0.46 &  0.0765 & 15 
\end{tabular}\label{eq:proparampositrons}\eeq

Finally, the source term due to DM DM annihilations in each point of the halo with DM density $\rho(\vec x)$ is
\beq \label{eq:Q}
Q = \frac{1}{2} \left(\frac{\rho}{M_{\rm DM}}\right)^2 f_{\rm inj},\qquad f_{\rm inj} = \sum_{k} \langle \sigma v\rangle_k \frac{dN_{e^+}^k}{dE}\eeq
where $k$ runs over all the channels with positrons in the final state, with the respective thermal averaged cross sections $\sigma v$.

One assumes steady state conditions, so that the first term of eq.\eq{diffeq} vanishes, and the solution for the positron flux at Earth can be written in a useful semi-analytical form~\cite{FornengoDec2007,HisanoAntiparticles}:
\beq 
\Phi_{e^+}(E,\vec r_{\odot}) = B \frac{v_{e^+}}{4\pi b(E)}
\frac{1}{2} \left(\frac{\rho_\odot}{M_{\rm DM}}\right)^2  \int_{E}^{M_{\rm DM}}  dE'~f_{\rm inj}(E')\cdot  I \left(\lambda_D(E,E')\right)
\label{eq:fluxpositrons}
\eeq
where $B\ge 1$ is an overall boost factor discussed below, $\lambda_D(E,E')$ is the diffusion length from energy $E'$ to energy $E$:
\beq \lambda_D^2= 4K_0 \tau_E \left[\frac{(E/\GeV)^{\delta-1} - (E'/\GeV)^{\delta-1}}{\delta-1}\right]
\label{eq:lambdaD}\eeq
and the adimensional `halo function' $I(\lambda_D)$~\cite{FornengoDec2007} fully encodes the galactic astrophysics and is independent on the particle physics model.
Its possible shapes are plotted in fig.\fig{HaloI} for the set of DM density profiles and positron propagation parameters that we consider.\footnote{
Formally, one finds that 
\beq
I(\lambda_D) = \sum_{n,m=1}^{\infty} J_0( \zeta_n r_\odot/ R )\ \sin (m \pi / 2) \ {\rm exp}\left[ -\left( \left( \frac{m \pi}{2 L} \right)^2 + \left(\frac{\zeta_n}{R}\right)^2 \right) \frac{\lambda_D^2}{4} \right] R_{n,m}
\eeq
where $J_i$ is the Bessel function of the first kind (cylindrical harmonic) of order $i$, $\zeta_n$ is the $n$-th zero of the $i=0$  function and $R_{n,m}$ corresponds to the Bessel- and Fourier-transform of $(\rho/\rho_\odot)^2$:
\beq
R_{n,m} = \frac{2}{J_1(\zeta_n)^2 R^2} \int_0^Rdr\, r J_0( \zeta_n r/ R ) \frac{1}{L} \int_{-L}^{+L} dz\, \sin(m \pi z / 2L) \left( \frac{\rho(r,z)}{\rho_\odot} \right)^2.
\eeq}
From the numerical computation we find that $I(\lambda_D)$ is well reproduced with a na\"ive fit function of the form 
\beq
I(\lambda_D) = a_0 + a_1 \tanh\left(\frac{b_1-\ell}{c_1}\right) \left[ a_2 \exp \left( -\frac{(\ell - b_2)^2}{c_2}\right) + a_3 \right]
\label{eq:fitpositrons}
\eeq
with $\ell = \log_{10}\lambda_D/_{\rm kpc}$ and the coefficients reported in table~\ref{table:fitpositrons}.

\begin{table}
$$
\begin{tabular}{c | l | rrrrrrrr}
\rm{Halo model} & \rm{Propagation} & $a_0$ & $a_1$ & $a_2$ & $a_3$ & $b_1$ & $b_2$ & $c_1$ & $c_2$\\
\hline
& \rm{min (M2)} & 0.500 & 0.774 & -0.448 & 0.649 & 0.096 & 192.8 & 0.211 & 33.88 \\
 \rm{NFW}   & \rm{med} & 0.502 & 0.621 & 0.688 & 0.806 & 0.891 & 0.721 & 0.143 & 0.071 \\
  & \rm{max (M1)} & 0.502 & 0.756 & 1.533 & 0.672 & 1.205 & 0.799 & 0.155 & 0.067 \\
  \hline
& \rm{min (M2)} & 0.500 & 0.791 & -0.448 & 0.636 & 0.096 & 192.8 & 0.211 & 33.86 \\
 \rm{Moore}   & \rm{med} & 0.503 & 0.826 & 0.938 & 0.610 & 0.912 & 0.762 & 0.162 & 0.055 \\
  & \rm{max (M1)} & 0.503 & 0.889 & 1.778 & 0.571 & 1.230 & 0.811 & 0.135 & 0.061 \\
  \hline
& \rm{min (M2)} & 0.500 & 0.903 & -0.449 & 0.557 & 0.096 & 192.8 & 0.210 & 33.91 \\
 \rm{isoT}   & \rm{med} & 0.495 & 0.629 & 0.137 & 0.784 & 0.766 & 0.550 & 0.193 & 0.296 \\
  & \rm{max (M1)} & 0.499 & 0.695 & 0.677 & 0.721 & 1.092 & 0.951 & 0.379 & 0.231
\end{tabular}
$$
\caption{\em Fit parameters for the expression in eq.~$(\ref{eq:fitpositrons})$ for the halo function $I(\lambda_D)$ that encodes the astrophysics of the production density and the propagation of positrons in the galactic halo.}
\label{table:fitpositrons}
\end{table}

The main features of the halo function can be understood as follows.
It is defined such that $I\simeq1$ at $\lambda_D \ll r_\odot,L$: all positrons created close enough to the Earth can reach it without loosing energy.
$I$ can exhibit a peak at $\lambda_D\sim r_\odot$ if positrons produced by DM DM annihilations around
 the galactic center are dominant and reach us after loosing some energy.
 If instead the diffusive region is thin we only receive positrons produced within
 a region $\sim L$ around the Earth: e.g.\ the dashed lines are for $L \sim1\,{\rm kpc}$.

\subsection{Results}
Unlike photons, where we look at the central cuspy region of the galactic DM halo
(such that the photon flux can be very large but also very uncertain), positrons do not have directionality.
Especially at energies just below the DM mass $M$, positrons are dominantly produced close to the solar system,
so that their flux is less affected by uncertainties in the DM profile.

However, the DM density in our galaxy might have local clumps that would enhance the positron flux by an unknown
`boost factor' $B\ge 1$. We take it as energy independent and with a value of $B = 10$. This is a simplifying (but widely used) assumption. Detailed recent studies~\cite{Lavalle1,Lavalle2,minispikes,Berez} find that a certain energy dependance can be present, subject to the precise choices of the astrophysical parameters. Within the uncertainty, these studies also converge towards small values of $B$ (except for extreme scenarios), with $B = 10$ still allowed.

\medskip

\begin{figure}[t]
\begin{center}
\includegraphics[width=0.95\textwidth]{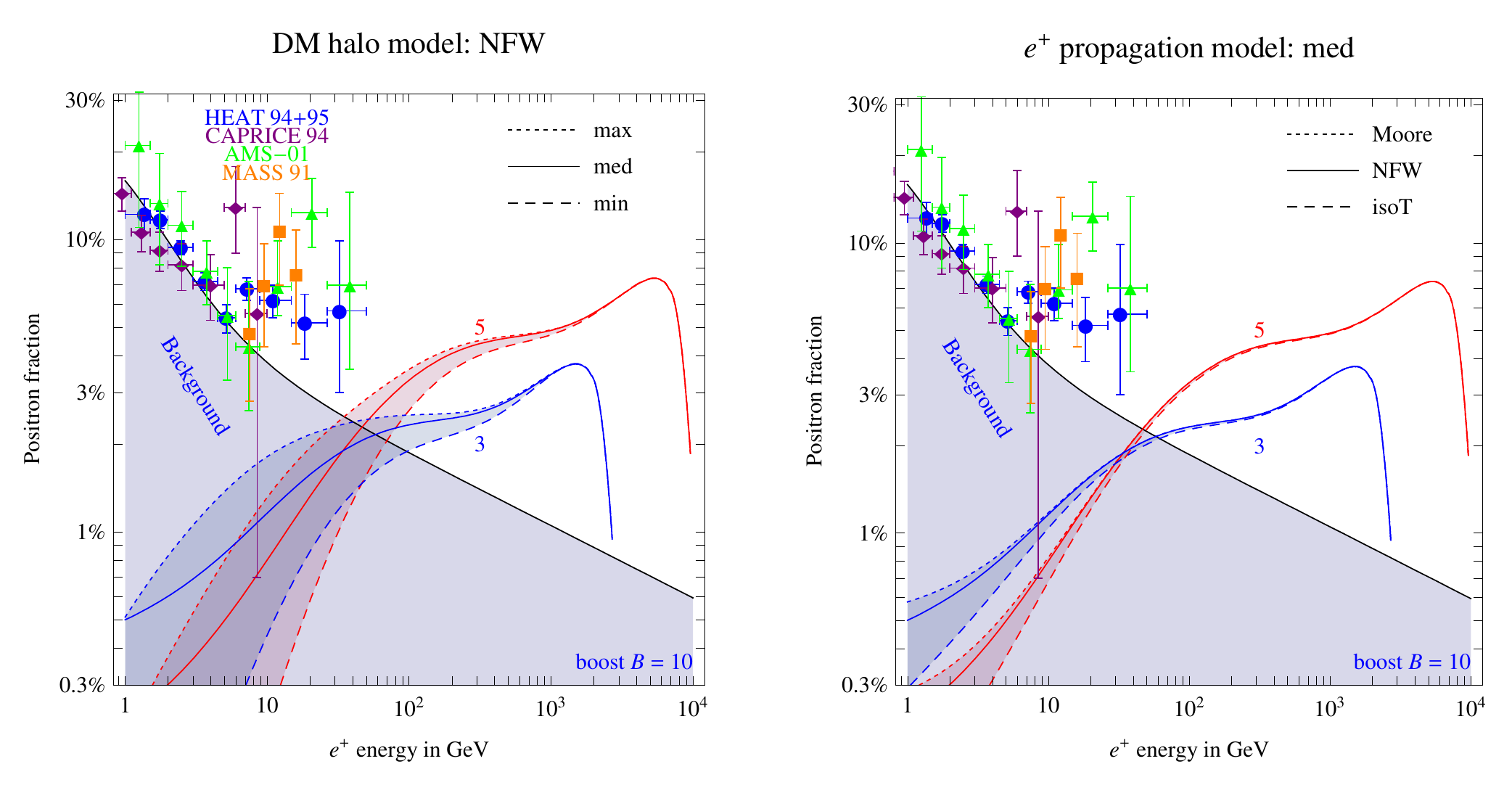}
\caption{\em\label{fig:PositronR} {\bf Positron fraction}, $N_{e^+}/(N_{e^+}+N_{e^-})$, generated by
$\DM \, \DM$ annihilations. The red (upper) curves refer to the 5-plet MDM candidate (eq.\ref{sys:sample}b). The blue (lower) ones to the 3-plet (eq.\ref{sys:sample}a).
In the left plot we fix the NFW halo profile and vary the $e^+$ propagation model.
In the right plot we fix the {\rm med} propagation model and vary the DM halo profile.
We assumed a boost factor $B=10$: notice that a signal above the background is present even for $B=1$, for the 5-plet case. The experimental data points are taken from~\cite{HEAT,CAPRICE,AMS01,MASS}.
}
\label{default}
\end{center}
\end{figure}

The results are shown in term of the energy spectrum of the positron flux at Earth from DM DM annihilations,
computed for several astrophysical models and compared with the expected background. The latter, believed to be mainly due to supernova explosions, is obtained by CR simulations~\cite{MoskalenkoStrong} and can be parameterized as described in~\cite{bkgpositrons} by $\Phi_{e^+}^{\rm bkg} = 4.5\, E^{0.7}/(1+650\, E^{2.3}+1500\, E^{4.2})$ for positron and 
$$\Phi_{e^-}^{\rm bkg} =\Phi_{e^-}^{\rm bkg,\, prim} +\Phi_{e^-}^{\rm bkg,\, sec} = 0.16\, E^{-1.1}/(1+11\, E^{0.9}+3.2\, E^{2.15}) + 0.70\, E^{0.7}/(1+110\, E^{1.5}+580\, E^{4.2})$$ for electrons, with $E$ always in units of GeV.
In fig.\fig{PositronR} we actually plot the positron fraction, $\Phi_{e^+}/(\Phi_{e^+}+\Phi_{e^-})$, as the flux $\Phi_{e^-}$ of cosmic ray electrons provides a convenient normalization, and the ratio does not depend on solar activity (see the discussion in the case of anti-protons).
We see that in the region at $E\sim M_\DM/3$ where the signal/background ratio
is maximal, the predicted signal is quite distinctive and does not significantly depend on unknown astrophysics, being manly generated by prompt positrons created close to the Earth.

The overall rate is however uncertain because the DM DM annihilation cross-sections  vary by about one order of magnitude within the narrow range of $M_\DM$ that reproduces the cosmological DM density
and because of the possible enhancement coming from the boost factor $B$.
We here assumed the sample values of eq.s~(\ref{sys:sample}) and a boost factor: $B=10$.
With this choice, the excess starts to appear just around the maximal energy probed by current experiments,
and would give a clear signal in the PAMELA experiment~\cite{PAMELA}.

A signal is still present even for $B=1$; however it appears only at higher energies, 
around the peaks of the signal curves in fig.\fig{PositronR} at $E\approx M_\DM/3\sim \TeV$.
This region of energies will be hopefully explored by the future AMS-02 experiment~\cite{AMS02}.\footnote{The experimental limitation on maximal energies arises because the energy is measured
from deflection of charged particles in the magnetic field of the spectrometer and above a certain threshold positrons are shadowed by the abundant spillover protons~\cite{spillover}.
The PAMELA experiment should soon release data about $e^+$ and $\bar p$ up to about 190 GeV and 270 GeV respectively, while AMS-02  might reach the TeV region.
}
Finally, we recall that the scalar triplet MDM predicts a signal about 16 times higher than the fermion triplet MDM.

\subsection{Synchrotron radiation}
Another possible DM signal is the synchrotron radiation from $e^\pm$ produced in DM DM annihilations.
For simplicity, since astrophysics is anyhow significantly uncertain,
we neglect the time and space dependence of $f$ in the diffusion equation\eq{diffeq}:
this amounts to assume that $e^\pm$ are trapped enough in the galaxy bulge that 
they loose there most of the energy, consequently maximizing the synchrotron signal.
Solving eq.\eq{diffeq} the $e^\pm$ energy spectrum is then given by
\beq f(E) =\frac{1}{b(E)}
\int_E^\infty dE' Q(E'),\qquad E= xM_{\rm DM},\eeq
and the energy spectrum of synchrotron radiation is then given by
\beq \frac{dP_\gamma}{dE_\gamma}\propto \frac{\sqrt{3}}{2\pi}\frac{e^3B}{m_e}
\int_0^1 dx~f(E=xM_{\rm DM}) F(r/x^2)\label{eq:fsyn}\eeq 
where $F(x)\equiv x \int_x^\infty K_{5/3}(\xi)d\xi \sim x^{1/3}e^{-x}$ is the synchrotron function.
The adimensional factor $r = 2 m_e^3 E_\gamma/3eBM^2_{\rm DM}$ encodes the dependence on $E_\gamma$, on the DM mass $M_{\rm DM}$ and
on astrophysics trough the uncertain magnetic field $B$. 
Numerically $r\approx 10^{-4}$ for a magnetic field
$B=\mu{\rm G}$, $M_{\rm DM}=1\TeV$ and $E_\gamma = 10 \,{\rm GHz}$.
The integral in eq.\eq{fsyn} is easily computed numerically; we here just notice that
one roughly has  $f(x) \propto 1/x^2$ due to the $E^2$ dependence of $b(E)$ and consequently
${dP_\gamma}/{dE_\gamma} \propto  E_\gamma^{-1/2}$.

WMAP observed of an apparent excess of radiowaves with $\nu \sim 20\,{\rm GHz}$ from the galactic center,
that might be due to synchrotron radiation from $e^\pm$ produced in DM DM annihilations~\cite{WMAPHaze}.
The angular dependence of the signal is precisely measured; however it depends on astrophysical issues: the DM density profile and
the $e^\pm$ propagation model. A cusped halo model allows to fit the anomaly.
The energy dependence of the signal has not been precisely measured and the MDM prediction is compatible with the WMAP haze.
Furthermore for the MDM values of $\sigma v$ and of $M_{\rm DM}$, and for reasonable values of the magnetic field $B$, the intensity is comparable
with the WMAP haze~\cite{WMAPHaze}.

\section{Antiprotons}

\subsection{Astrophysics}
The propagation of anti-protons through the galaxy is described by a diffusion equation analogous to the one for positrons.
Again, the number density of anti-protons per unit energy $f(t,\vec x,T) = dN_{\bar p}/dT$ vanishes on the surface of the cylinder at $z=\pm L$ and $r=R$. $T=E-m_p$ is the $\bar p$ kinetic energy, conveniently used instead of the total energy $E$ (a distinction which will not be particularly relevant for our purposes as we look at energies much larger than the proton mass $m_p$).
Since $m_p\gg m_e$ we can neglect the energy loss term, and the diffusion equation for $f$ is
\beq 
\label{eq:diffeqp}
\frac{\partial f}{\partial t} - K(T)\cdot \nabla^2f + \frac{\partial}{\partial z}\left( {\rm sign}(z)\, f\, V_{\rm conv} \right) = Q-2h\, \delta(z)\, \Gamma_{\rm ann} f  
\eeq
where:
\begin{itemize}
\item[-] The pure diffusion term can again be written as $K(T) = K_0 \beta \, (p/\GeV)^\delta$, where $p = (T^2 +2 m_p T)^{1/2}$ and 
$\beta = v_{\bar p}/c = \left(1-m_p^2/(T+m_p)^2\right)^{1/2}$ are the antiproton momentum and velocity. $\delta$ and $K_0$ are given in eq.\eq{proparam}.

\item[-] The $V_{\rm conv}$ term corresponds to a convective wind, assumed to be constant and directed outward from the galactic plane, that tends to push away $\bar p$ with energy $T \circa{<}10\, m_p$. Its value is given in eq.\eq{proparam}.

\item[-] The source term $Q$ due to DM DM annihilations has a form fully analogous to eq.\eq{Q}, with $E$ now formally replaced by $T$.

\item[-]
The last term in eq.\eq{diffeqp} describes the annihilations of $\bar p$ on interstellar protons in the galactic plane
(with a thickness of $h=0.1\,{\rm kpc} \ll L$) with rate
$\Gamma_{\rm ann} = (n_{\rm H} + 4^{2/3} n_{\rm He}) \sigma^{\rm ann}_{p\bar{p}} v_{\bar{p}}$,
where $n_{\rm H}\approx 1/{\rm cm}^3$ is the hydrogen density, $n_{\rm He}\approx 0.07\, n_{\rm H}$ is the Helium density (the factor $4^{2/3}$ accounting for the different geometrical cross section in an effective way)
and $ \sigma^{\rm ann}_{p\bar{p}}$ is given by \cite{crosssection,HisanoAntiparticles}
\beq
\sigma_{p \bar p}^{\rm ann} = \left\{ 
\begin{array}{ll}
661\, (1+0.0115\, T^{-0.774} - 0.984\, T^{0.0151})\ {\rm mbarn}, & {\rm for}\ T < 15.5\, {\rm GeV} \\ 
36\, T^{-0.5}\ {\rm mbarn}, & {\rm for}\ T \geq 15.5\, {\rm GeV}
\end{array}
 \right. .
\label{eq:sigmaann}
\eeq
\item[-]
We neglect the effect of ``tertiary anti-protons''. This refers to primary $\bar p$ after they have undergone non-annihilating interactions on the matter in the galactic disk, losing part of their energy. The effect can be included in terms of an absorption term analogous to the last term of eq.\eq{diffeqp} but proportional to a different $\sigma^{\rm non-ann}$, and of a re-injection term $Q^{\rm tert}$ proportional to the integrated cross section over $f(T)$. The full solution of the resulting integro-differential equation can be found in~\cite{DonatoApJ563}. The effect of tertiaries is mainly relevant at low energies $T \lesssim$ few GeV.
 \end{itemize}

The set of propagation parameters in the case for anti-protons that we adopt has been deduced in \cite{DonatoPRD69} from a variety of cosmic ray data and modelization (see \cite{MaurinApJ555}): 
\beq \begin{tabular}{ccccc}
Model  & $\delta$ & $K_0$ in kpc$^2$/Myr & $L$ in kpc & $V_{\rm conv}$ in km/s \\
\hline 
min  & 0.85 &  0.0016 & 1  &13.5\\
med  & 0.70 &  0.0112 & 4  & 12 \\
max  & 0.46 &  0.0765 & 15 &5
\end{tabular}\label{eq:proparam}\eeq

Assuming steady state conditions the first term in the diffusion equation vanishes, and the equation can be solved analytically~\cite{methodPbar, TailletRRDA, MaurinApJ555}. In the ``no-tertiaries" approximation that we adopt, the solution for the antiproton flux at the position of the Earth $ \Phi_{\bar p}(T,\vec r_\odot) = v_{\bar p}/(4\pi) f $ acquires a simple factorized form (see e.g.~\cite{DonatoPRD69})
\beq
\Phi_{\bar p}(T,\vec r_\odot) = B \frac{v_{\bar p}}{4\pi}  \left(\frac{\rho_\odot}{M_{\rm DM}}\right)^2 R(T)   \sum_k \frac{1}{2} \langle \sigma v\rangle_k \frac{dN^k_{\bar p}}{dT}
\label{eq:RT}
\eeq
where $B$ is the boost factor. The $k$ index runs over all the annihilation channels with anti-protons in the final state, with the respective cross sections; this part contains the particle physics input. The function $R(T)$ encodes all the astrophysics and is plotted in fig.\fig{HaloI} for various halo and propagation models.\footnote{Formally, it is given by
\beq
R(T) = \sum_{n=1}^\infty J_0\left(\zeta_n \frac{r_\odot}{R}\right) 
{\rm exp}\left[ -\frac{V_{\rm conv} L}{2 K(T)} \right]
\frac{y_n(L)}{A_n \sinh(S_n L/2)}
\label{eq:R}
\eeq
with 
\beq
y_n(Z) = \frac{4}{J_1^2(\zeta_n) R^2} \int_0^R dr\, r \, J_0( \zeta_n r/ R ) 
\int_0^Z dz \, {\rm exp}\left[ \frac{V_{\rm conv} (Z-z)}{2 K(T)} \right] {\rm sinh}\left(S_n (Z-z)/2\right)
\left( \frac{\rho(r,z)}{\rho_\odot} \right)^2 
\label{eq:y}
\eeq
The coefficients $A_n = 2 h \Gamma_{\rm ann} + V_{\rm conv} + K(T)\, S_n\coth(S_n L/2)$ with $S_n=\left(V_{\rm conv}^2/K(T)^2 + 4\zeta_n^2/R^2\right)^{1/2}$ encode the effects of diffusion.}
From the numerical computation we find that $R(T)$ is well reproduced with a fit function of the form 
\beq
{\rm log}_{10}\left[R(T)/{\rm Myr}\right] = a_0 + a_1\, \tau + a_2\,  \tau^2 + a_3\,  \tau^3 + a_4\,  \tau^4
\label{eq:fitantiprotons}
\eeq
with $\tau = \log_{10}T/\GeV$ and the coefficients reported in table~\ref{table:fitantiprotons}.

\begin{table}
$$
\begin{tabular}{c | c | r r r r r}
\rm{Halo model} & \rm{Propagation} & $a_0$ & $a_1$ & $a_2$ & $a_3$ & $a_4$ \\[0.3mm]
\hline
& \rm{min} & 0.913 & 0.601 & -0.309 & -0.036 & 0.0122 \\
 \rm{NFW} 	 & \rm{med} & 1.860 & 0.517 & -0.293 & -0.0089 & 0.0070 \\
	 & \rm{max} & 2.740 & -0.127 & -0.113 & 0.0169 & -0.0009 \\
\hline
 & \rm{min} & 0.894 & 0.606 & -0.299 & -0.041 & 0.0128 \\
 \rm{Moore} 	 & \rm{med} & 1.870 & 0.553 & -0.289 & -0.0149 & 0.0079 \\
	 & \rm{max} & 2.810 & -0.119 & -0.117 & 0.0181 & -0.0010 \\
\hline
& \rm{min} & 0.927 & 0.590 & -0.315 & -0.0319 & 0.0115 \\
 \rm{isoT}  	 & \rm{med} & 1.790 & 0.399 & -0.315 & 0.0162 & 0.0031 \\
 	 & \rm{max} & 2.480 & -0.156 & -0.098 & 0.0132 & -0.0005
\end{tabular}
$$
\caption{\em Fit parameters for the expression in eq.~$(\ref{eq:fitantiprotons})$ for the propagation function $R(T)$ that encodes the astrophysics of the production density and the propagation of antiprotons in the galactic halo.}
\label{table:fitantiprotons}
\end{table}

Finally, for completeness we also take into account the average solar modulation effect, although it is relevant only for non-relativistic $\bar p$: the solar wind decreases the kinetic energy $T$ and momentum $p$
of charged cosmic rays such that the energy spectrum $d\Phi_{\bar{p}\oplus}/dT_\oplus$
of anti-protons that reach the Earth with energy $T_\oplus$ and momentum $p_\oplus$
 is approximatively related to their energy spectrum in the interstellar medium, $d\Phi_{\bar p}/dT$,
as~\cite{GA}
\beq \frac{d\Phi_{{\bar p}\oplus}}{dT_\oplus} = \frac{p_\oplus^2}{p^2} \frac{d\Phi_{\bar p}}{dT},\qquad
T= T_\oplus + |Ze| \phi_F, \qquad
p^2 = 2m_p T+T^2.
\eeq
The so called Fisk potential $\phi_F$ parameterizes in this effective formalism the kinetic energy loss. A value of $\phi_F= 0.5\, {\rm GV}$ is characteristic of a minimum of the solar cyclic activity, corresponding to the period in which most of the observations have been done in the second half of the 90's and are being done now.

\begin{figure}[t]
\begin{center}
\includegraphics[width=0.45\textwidth]{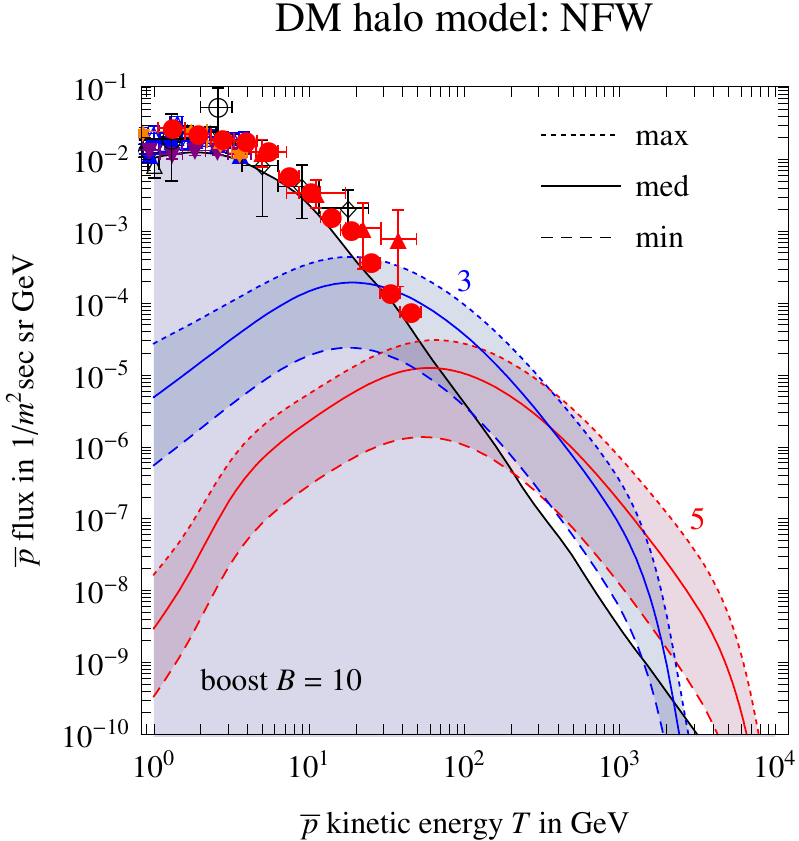}\qquad
\includegraphics[width=0.45\textwidth]{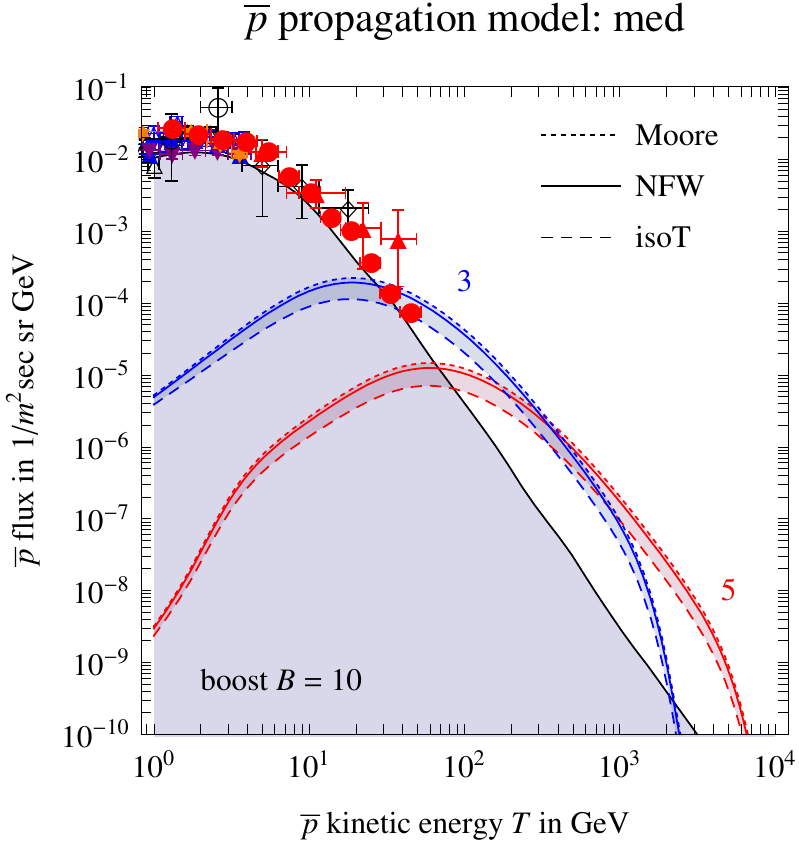}
\caption{\em \label{fig:Rp} The {\bf antiproton flux} generated by {\rm DM DM} annihilations for the case of the fermion 3-plet (blue, higher) and 5-plet (red, lower) MDM candidates, compared with the astrophysical $\bar p$ background (shaded area) and experimental data.
In the left plot we fix the NFW halo profile and vary the $\bar p$ propagation model. 
In the right plot we fix the {\rm med} propagation model and vary the DM density profile.
The compilation of data point includes results from the {\sc BESS}~\cite{BESS}, {\sc MASS}~\cite{MASSP}, {\sc CAPRICE}~\cite{CAPRICEP} and {\sc AMS-01}~\cite{AMS01P} experiments, as well as the preliminary results from the PAMELA experiment~\cite{PAMELApreliminary}.
A boost factor $B=10$ is assumed, but a signal is present even for $B=1$.}
\end{center}
\end{figure}

\subsection{Results}
Fig.\fig{Rp} shows the results for final $\bar p$ flux at earth (at the top of the atmosphere) from DM DM annihilations, compared to the background and to the currently available experimental data. The background is borrowed from the detailed analysis in~\cite{BringmannSalati}, the results of which we find to be well reproduced by a fitting function of the form 
$${\rm log}_{10} \Phi_{\bar p}^{\rm bkg} = -1.64 + 0.07\, \tau - \tau^2 - 0.02\, \tau^3 + 0.028\, \tau^4$$
with $\tau = {\rm log}_{10} T/\GeV$. We take for definiteness the flux corresponding to the `med' propagation parameters; see~\cite{BringmannSalati} for a complete discussion on the effects of changing that. Particularly favorable is the fact that the uncertainty in the estimates of the background is quite narrow around $10 - 100$ GeV, where results are expected soon.

The shape of the spectrum appears to be relatively independent from the propagation model (fig.\fig{Rp}a) and the halo profile (fig.\fig{Rp}b).
Different $\bar p$ propagation models instead change the overall signal rate by about one orders of magnitude, 
consistently with previous results in the literature~\cite{BringmannSalati, DonatoApJ563, DonatoPRD69}.
Different halo profiles with fixed $\rho_\odot$
make only a difference of a factor of a few, which can be interpreted in terms of the fact that the signal is not dominated by the far galactic center region, where profiles differ the most.

As for the case of the positrons, we have plotted the results assuming a modest and energy independent boost factor $B=10$ (in principle this boost factor and its properties are different from those for positrons~\cite{Lavalle2}). In this case the excesses appear in the range of energies soon to be explored. Even for a boost factor $B=1$ a signal is present above the background for most choices of parameters, although it would show at higher energies.
Again the scalar triplet predicts a signal about 16 times larger than the fermion triplet.

\section{Conclusions}\label{concl}
We computed the  indirect detection signatures (fluxes of positrons, anti-protons, $\gamma$ and synchrotron radiation) 
as predicted by Minimal Dark Matter.
We focussed on three particularly interesting MDM candidates: the automatically stable ermion 5-plet with hypercharge $Y=0$; the wino-like fermion 3-plet with $Y=0$; its scalar analogous.
We fixed the MDM masses and annihilation cross sections to the central values predicted in terms of the measured cosmological abundance. These values are listed in eq.~(\ref{sys:sample}).
Since MDM predicts multi-TeV masses and the Sommerfeld electroweak non-perturbative enhancement of the $\DM \, \DM$ annihilation cross sections into $W^+W^-$, $\gamma\gamma$, $\gamma Z$, $ZZ$, the signals for indirect detection turn out to be distinctive, reaching multi-TeV energies and being above the astrophysical background. 
The spectral shapes are characteristic of DM DM annihilations into SM vectors.
\medskip

We recomputed independently most of the ingredients that are necessary for the analysis, finding agreement with results  in the literature when these are available. The spectra at production were recomputed taking into account spin correlations among SM vectors,
by implementing the MDM interactions in {\sc MadGraph}~\cite{MadGraph} or by a custom-built MC routine and hadronizing the resulting MonteCarlo events with {\sc Pythia}~\cite{Pythia}. The propagation diffusion-loss equation for positrons and anti-protons was solved following the semi-analytic prescriptions discussed in the literature and for a variety of propagation models and halo profiles. 
We offer plots (fig.\fig{Fragmentation} and \fig{HaloI}) and simple fit functions for all these ingredients.

\medskip

Fig.\fig{gamma} shows the predicted {\bf photon} energy spectrum.


\medskip

The {\bf positron} flux is shown in fig.~\fig{PositronR}. This signal is only very mildly affected by the DM density profile, except for multiplicative uncertainties due to the uncertain local DM density (we assumed $\rho_\odot=0.3\GeV/{\rm cm}^3$) and due to the boost factor $B\ge1$, taken for simplicity as energy independent.
On the other hand, $e^+$ fluxes somewhat depend on the $e^+$ propagation model in our galaxy;
this uncertainty  will be reduced by future measurements of cosmic rays and is present only at $E\ll M_\DM$.
Indeed we do not know if positrons produced around the galactic bulk escape from the galaxy
or reach us after loosing most of their energy (giving also an interesting synchrotron radiation signal).
On the contrary positrons produced in the region of the galaxy close to the solar system surely reach us without loosing
significant energy: such positrons with $E\circa{<} M_\DM$ give a detectable MDM signal even if $B=1$.

\medskip

The {\bf anti-proton} flux is shown in fig.\fig{Rp}. Different  $\bar p$ propagation models give $\bar p$ fluxes that differ by about one order of magnitude, and different DM density profiles give fluxes that differ by about a factor 2. A detectable signal is again typical, and its energy spectrum (as predicted by MDM) is not significantly affected by astrophysical uncertainties. 

\medskip

Experimental results are expected soon.


\paragraph{Acknowledgements}
We thank Johan Alwall, Gianfranco Bertone, Torsten Bringmann, Nicolao Fornengo, Fabio Maltoni, Michelangelo Mangano, Stephen Mrenna, Torbj\"orn Sj\"ostrand, Mike Capell (AMS-02 collaboration) and Vitaly Choutko (AMS-02 collaboration). We also thank particularly Pierre Brun and Julien Lavalle for many useful suggestions and discussions. We thank Alejandro Ibarra and David Tran for spotting some typos in the first version of the manuscript.  M.C. acknowledges support from INFN under the postdoctoral grant 11067/05 at the early stages of this work.

\bigskip

\appendix

\footnotesize

\begin{multicols}{2}
  
\end{multicols}

\normalsize

\newpage

\section*{Addendum: preliminary PAMELA results}\label{in}
\setcounter{equation}{0}
\renewcommand{\theequation}{\arabic{equation}A}

\begin{figure}
$$\includegraphics[width=0.9\textwidth]{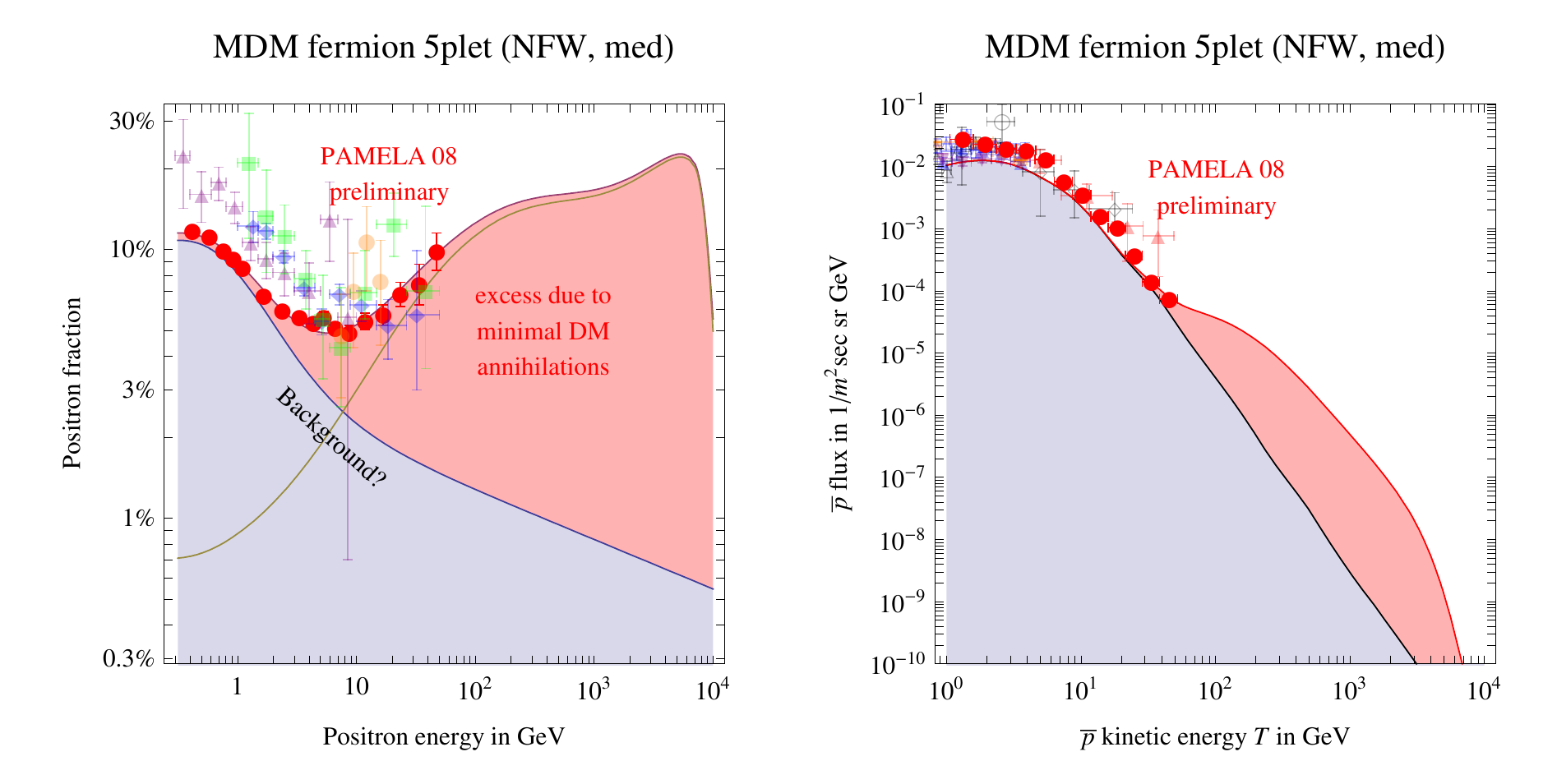}$$
\caption{\label{fig:PAM}\em Predicions of the Minimal Dark Matter fermion quintuplet compared to
preliminary PAMELA data.  
In both figures the boost factor equals
$B=3$ if $\langle\sigma v \rangle_{WW}=9\cdot 10^{-23}\,{\rm cm}^3/{\rm sec}$ which corresponds to $M=9.2\TeV$,  
or $B=30$ if $\langle\sigma v \rangle_{WW}=1.1\cdot 10^{-23}\,{\rm cm}^3/{\rm sec}$, obtained for $M=9.6\TeV$,
within the range inferred from the measured DM cosmological abundance.
}\end{figure}

The PAMELA collaboration presented preliminary data about the spectra of
anti-proton and positron cosmic rays.
No anomaly seems present in the anti-proton data, see fig.\fig{PAM}b~\cite{PAM}.
Compatibly with measurements from previous experiments,
an anomaly possibly due to DM annihilations seems present in the data about the $e^+/(e^++e^-)$ fraction,
as indicated by the fact that the positron fraction grows at high energy, see fig.\fig{PAM}a~\cite{PAM}.
At energies below 10 GeV PAMELA measures a  flux somewhat lower than previous experiments: 
this seems due to the variation in solar activity~\cite{PAM}.
Previous experimental results have significantly larger uncertainties and can be here ignored.

\medskip

We here compare these PAMELA preliminary data with the univocal predictions of 
the automatically stable Minimal Dark Matter candidate: the fermion quintuplet.
Its contribution was predicted in the present paper in figures 5 and 6.
In fig.\fig{PAM} we sum it to
the expected astrophysical background, assuming the standard NFW halo profile and the `medium'
model for positron diffusion in our galaxy.

We see that the anomaly in positrons can be well reproduced;
thanks to the Sommerfeld enhancement of the DM DM annihilation cross section
in the non-relativistic limit, a modest
boost factor is needed, $B=30$ for the central value of $M=9.6\TeV$ suggested by the cosmological DM abundance, 
and $3<B<100$ in the $3\sigma$ range of $M$ compatible with cosmology,
$9.2\TeV<M<10.2\TeV$.

In fig\fig{PAM}b we show the corresponding effect in the anti-proton flux.
Although other possibilities exist, for  simplicity we assumed the same energy-independent boost factor
as for positrons.
In such a case, an anomaly in anti-protons is expected at higher energies, compatibly with present PAMELA data.
In the future, PAMELA should provide experimental data up to maybe 200 GeV for positrons, and 100 GeV for anti-protons.
Higher energy will be reached by the future AMS experiment.
These measurements will test the predictions illustrated in fig.\fig{PAM}.

\medskip

To illustrate the discriminatory power of such future data, we point out
an alternative interpretation of the PAMELA anomaly.
In order to minimize the needed  boost factor,
we consider an hypothetical ad hoc DM candidate with mass $M$ just above $60\GeV$ that annihilates directly into
$ e^-e^+$ with $100\%$ branching ratio.
Assuming negligible co-annihilations and that 
DM DM annihilations are $s$-wave dominated, the cosmological abundance is
reproduced for $\langle\sigma v\rangle \approx 3\cdot 10^{-26}\,{\rm cm}^3/{\rm sec}$. Inserting $dN_{e^+}/dE _{e^+}= \delta(E_{e^+}-M)$ in eq.\eq{Q},
and taking into account positron energy loss in our galaxy as in eq.\eq{fluxpositrons},
we find that a boost factor $B\approx 10$ gives a positron fraction that  somewhat resembles the PAMELA anomaly.
If this other interpretation is true, PAMELA will see that the positron excess abruptly terminates at $E_{e^+}>M$.

\footnotesize
\small



\begin{thebibliography}{nn}


\bibitem{MDM}
  \art[hep-ph/0512090]{M. Cirelli, N. Fornengo, A. Strumia}{Nucl. Phys.}{B753}{178}{2006}.


\bibitem{MDM2}
\art[0706.4071]{M.~Cirelli, A.~Strumia and M.~Tamburini}{Nucl.\ Phys.}{B787}{152}{2007}.


\bibitem{cosmoDM}
These numbers summarize 
various recent global analyses of cosmological data
within the $\Lambda$CDM model
that found compatible values and uncertainties:
\hepart[astro-ph/0603449]{D. N. Spergel {\it et al.} [WMAP collaboration]},
\art[astro-ph/0607086]{M. Cirelli and A. Strumia}{JCAP}{0612}{013}{2006},
\art[astro-ph/0608632]{M. Tegmark et al.}{Phys. Rev.}{D74}{123507}{2006}.


\bibitem{pioneers}
\art{J. Silk and M. Srednicki}{Phys. Rev. Letters}{53}{624}{1984}
\art{J.~R.~Ellis, R.~A.~Flores, K.~Freese, S.~Ritz, D.~Seckel and J.~Silk}{Phys.\ Lett.}{B 214}{403}{1988}.
See also the pioneering work in 
\art{J. E. Gunn, B.W. Lee, I. Lerche, D. N. Schramm and G. Steigman}{Astrophys. Journ.}{223}{1015-1031}{1978}.
\art{F. W. Stecker}{Astrophys. Journ.}{223}{1032-1036}{1978}.
\art{Ya. B. Zeldovich, A. A. Klypin, M. Yu. Khlopov and V. M. Chechetkin}{Yadernaya
Fizika}{31}{1286-1294}{1980}. [English translation: Sov. J. Nucl. Phys. (1980) V.31, PP. 664-669]

\bibitem{PAMELA}
\art[astro-ph/0608697]{P.~Picozza {\it et al.}}{Astropart.\ Phys.}{27}{296}{2007}.
Web page: \myurl{pamela.roma2.infn.it/index.php}{pamela.roma2.infn.it/index.php}.

\bibitem{AMS02}
Web page: \myurl{ams.cern.ch}{ams.cern.ch}.

  
  \bibitem{Hisano}
\art[hep-ph/0407168]{J.~Hisano, S.~Matsumoto, M.~M.~Nojiri and O.~Saito}{Phys.\ Rev.\ D}{71}{015007}{2005}.
\art[hep-ph/0212022]{J.~Hisano, S.~Matsumoto and M.~M.~Nojiri}{Phys.\ Rev.\ D}{67}{075014}{2003}. 
\art[hep-ph/0307216]{J.~Hisano, S.~Matsumoto and M.~M.~Nojiri}{Phys.\ Rev.\ Lett.}{92}{031303}{2004}.
\art[hep-ph/0412403]{J.~Hisano, S.~Matsumoto, M.~M.~Nojiri and O.~Saito}{Phys.\ Rev.\ D}{71}{063528}{2005}.
\art[hep-ph/0511118]{J. Hisano, S. Matsumoto, O. Saito, M. Senami}{Phys. Rev.}{D73}{055004}{2006}.


\bibitem{Barger}
\art[hep-ph/9301265]{D. Chang and W.-Y. Keung}{Phys. Lett.}{B305}{1993}{261}.
\art[0708.1325]{V.~Barger, W.~Y.~Keung, G.~Shaughnessy and A.~Tregre}{Phys.\ Rev.\  D}{76}{095008}{2007}. 


\bibitem{MadGraph}
\art[hep-ph/0208156]{F.~Maltoni and T.~Stelzer}{JHEP}{0302}{2003}{027}.
\art[hep-ph/9401258]{T.~Stelzer and W.~F.~Long}{Comput.\ Phys.\ Commun.}{81}{1994}{357}.
\art[0706.2334]{J.~Alwall {\it et al.}}{JHEP}{0709}{2007}{028}.
Web page: \myurl{madgraph.phys.ucl.ac.be}{madgraph.phys.ucl.ac.be}.



\bibitem{Pythia}
T. Sj\"ostrand, S. Mrenna and P. Skands,
to appear in Comput. Phys. Comm. [0710.3820],
JHEP05 (2006) 026.
Web page: \myurl{www.thep.lu.se/~torbjorn/Pythia.html}{www.thep.lu.se/$\sim$torbjorn/Pythia.html}


\bibitem{BergstromBringmann}
\art[hep-ph/0507229]{L.~Bergstrom, T.~Bringmann, M.~Eriksson and M.~Gustafsson}{Phys.\ Rev.\ Lett.}{\bf 95}{241301}{2005}.
\art[0710.3169]{T.~Bringmann, L.~Bergstrom and J.~Edsjo}{JHEP}{0801}{2008}{049}.

\bibitem{IsoT} 
\art{J. N. Bahcall and R. M. Soneira}{Astrophys. J. Suppl.}{44}{73}{1980}.


\bibitem{NFW}
\art[astro-ph/9611107]{J. Navarro, C. Frenk, S. White}{Astrophys. J.}{490}{493}{1997}.


\bibitem{Moore04}
\art[astro-ph/0402267]{J.~Diemand, B.~Moore and J.~Stadel}{Mon.\ Not.\ Roy.\ Astron.\ Soc.}{353}{624}{2004}.

\bibitem{Kamionkowski}
See for instance the recent analysis in: 
  \hepart[0801.3269]{M.~Kamionkowski and S.~M.~Koushiappas}.
  
  
\bibitem{smoothing}
\art[astro-ph/0506389]{A.~Barrau, P.~Salati, G.~Servant, F.~Donato, J.~Grain, D.~Maurin and R.~Taillet}{Phys.\ Rev.}{D72}{063507}{2005}. 
  
  \bibitem{HESS} 
Web page of the H.E.S.S.\ project:
\myurl{www.mpi-hd.mpg.de/hfm/HESS}{www.mpi-hd.mpg.de/hfm/HESS}.
\art[astro-ph/0603021]{H.E.S.S. collaboration}{Nature}{439}{695}{2006}.



\bibitem{DiffusionCylinder}
The now standard ``two-zone diffusion model" introduced in 
\art{V.L. Ginzburg, Ya.M. Khazan, V.S. Ptuskin}{Astrophysics and Space Science}{68}{295-314}{1980},  
\art{W.R. Webber, M.A. Lee, M. Gupta}{Astrophysical Journal}{390}{96-104}{1992}.


\bibitem{FornengoDec2007}
\hepart[0712.2312]{T.~Delahaye, R.~Lineros, F.~Donato, N.~Fornengo and P.~Salati}.

\bibitem{HisanoAntiparticles}
\art[hep-ph/0511118]{J. Hisano, S. Matsumoto, O. Saito, M. Senami}{Phys. \
Rev.}{D73}{055004}{2006}.

\bibitem{Berez}
\art[astro-ph/0301551]{V.~Berezinsky, V.~Dokuchaev and Y.~Eroshenko}{Phys.\ Rev.}{D68}{103003}{2003}. 

\bibitem{Lavalle1}
\hepart[astro-ph/0603796]{J.~Lavalle, J.~Pochon, P.~Salati and R.~Taillet}

\bibitem{Lavalle2}
\hepart[0709.3634]{J.~Lavalle, Q.~Yuan, D.~Maurin and X.~J.~Bi}.

\bibitem{minispikes}
\art[0704.2543]{P.~Brun, G.~Bertone, J.~Lavalle, P.~Salati and R.~Taillet}{Phys.\ Rev.}{D76}{083506}{2007}. 
    
  
\bibitem{HEAT}
\art[astro-ph/9703192]{HEAT Collaboration:  S.~W.~Barwick {\it et al.}}{Astrophys.\ J.}{482}{L191}{1997}.
  
  \bibitem{CAPRICE}
CAPRICE Collaboration: \art{M. Boezio, P. Carlson, T. Francke, and N. Weber et al.}{The Astrophysical Journal}{532}{653-669}{2000}. Web page: \myurl{www.roma2.infn.it/research/comm2/caprice}{www.roma2.infn.it/research/comm2/caprice}
  
  
    \bibitem{AMS01}
  AMS-01 Collaboration: \art[astro-ph/0703154]{M. Aguilar et al.}{Phys. Lett.}{B646}{145-154}{2007}
  Web page: \myurl{ams.cern.ch/AMS/ams01_homepage.html}{ams.cern.ch/AMS/ams01\_homepage.html}


\bibitem{MASS}
\art{C. Grimani et al.}{Astron. Astrophys.}{392}{287-294}{2002}.



\bibitem{MoskalenkoStrong}
\art[astro-ph/9710124]{I.~V.~Moskalenko and A.~W.~Strong}{Astrophys.\ J.}{493}{694}{1998}.

\bibitem{bkgpositrons}
\art[astro-ph/9808243]{E.~A.~Baltz and J.~Edsjo}{Phys.\ Rev.}{D59}{023511}{1999}.

\bibitem{spillover}
See e.g.~\cite{PAMELA} and \hepart[0710.2458]{P.~Brun}.
  
    \bibitem{WMAPHaze}
  \hepart[astro-ph/0409027]{D.~P.~Finkbeiner}.
\art[0705.3655]{D.~Hooper, D.~P.~Finkbeiner and G.~Dobler}{Phys.\ Rev.}{D 76}{083012}{2007}.


  
\bibitem{crosssection}
\art{L.~C.~Tan and L.~K.~Ng}{J.\ Phys.}{G 9}{227}{1983}.


\bibitem{DonatoApJ563}
\art[astro-ph/0103150]{F.~Donato, D.~Maurin, P.~Salati, A.~Barrau, G.~Boudoul and R.~Taillet}{Astrophys.\ J.}{563}{172}{2001}.
  
  \bibitem{DonatoPRD69}
 \art[astro-ph/0306207]{F.~Donato, N.~Fornengo, D.~Maurin and P.~Salati}{Phys.\ Rev.}{D 69}{ 063501}{2004}.
  

  \bibitem{MaurinApJ555}
\art[astro-ph/0101231]{D.~Maurin, F.~Donato, R.~Taillet and P.~Salati}{Astrophys.\ J.}{555}{585}{2001}.
  
\bibitem{methodPbar}
\art[astro-ph/9606174]{P.~Chardonnet, G.~Mignola, P.~Salati and R.~Taillet}{Phys.\ Lett.}{B 384}{161}{1996}.
\art[astro-ph/9804137]{A.~Bottino, F.~Donato, N.~Fornengo and P.~Salati}{Phys.\ Rev.}{D 58}{123503}{1998}.
  See also: 
\art[astro-ph/9902012]{L.~Bergstrom, J.~Edsjo and P.~Ullio}{Astrophys.\ J.}{526}{215}{1999}.


\bibitem{TailletRRDA}  
\hepart[astro-ph/0212111]{D.~Maurin, R.~Taillet, F.~Donato, P.~Salati, A.~Barrau and G.~Boudoul}.


\bibitem{GA}
\art{L.J. Gleeson and W.I. Axford}{ApJ}{149}{L115}{1967} and 
\art{L.J. Gleeson and W.I. Axford}{ApJ}{154}{1011}{1968}.


\bibitem{BESS}
BESS 95+97: \art[astro-ph/9906426]{S. Orito et al.}{Phys. Rev. Lett.}{84}{1078-1081}{2000}. 
BESS 98: \art[astro-ph/0010381]{T. Maeno et al.}{Astropart. Phys.}{16}{121-128}{2001}.
BESS 2000: \art[astro-ph/0109007]{Y. Asaoka et al.}{Phys. Rev. Lett.}{88}{051101}{2002}.

\bibitem{MASSP}
MASS 91: G. Basini et al., Proc. $26^{\rm th}$ International Cosmic Ray Conference, Salt Lake City, USA(1999), OG.1.1.21.

\bibitem{CAPRICEP}
CAPRICE 94: \art{M. Boezio et al.}{Astrophys. J.}{487}{415-423}{1997}.
CAPRICE 98: \art[astro-ph/0103513]{M. Boezio et al.}{Astrophys. Jour.}{561}{787}{2001}.

\bibitem{AMS01P}
AMS-01: \art{M. Aguilar et al.}{Phys. Rept.}{366}{331-405}{2002}, 
Erratum-ibid. 380, 97-98 (2003). Data points: Vitaly Choutko, private communication.

\bibitem{PAMELApreliminary}
See e.g. P. Picozza, presentation at the {\it Eight UCLA Symposium: Sources and Detection of Dark Matter and Dark Energy in the Universe}, February 20-22, 2008, Marina del Rey, California, USA. We make use of the measured proton flux as well. 

\bibitem{BringmannSalati}
\art[astro-ph/0612514]{T.~Bringmann and P.~Salati}{Phys.\ Rev.}{D 75}{083006}{2007}.


\end{thebibliography}

\begin{thebibliography}{A00}
 \setcounter{enumiii}{0}
\renewcommand{\theenumiii}{\arabic{enumiii}}
\newcommand{\mybibitem}[1]{\addtocounter{enumiii}{1}\bibitem[\theenumiii A]{#1}}

\mybibitem{PAM} 
The data points shown in our figure have been graphically extracted from a
photo of a slide shown by M. Boezio at the IDM08 conference, Stochkolm, 20/08/2008.


\end{thebibliography}
\end{document}